\documentclass{aa}
\usepackage{graphics}
\input psfig.sty

\newcommand{\ltsima} {$\; \buildrel < \over \sim \;$}
\newcommand{\lta} {\lower.5ex\hbox{\ltsima}}

\begin{document}

\setcounter{table}{0}
\thesaurus{11.1.2,11.5.1,11.10.1,11.14.1}
\title{The HST snapshot survey of the B2 sample of low 
luminosity radio-galaxies: a picture gallery 
\thanks{Based on observations with the NASA/ESA Hubble Space Telescope, 
obtained at the
Space Telescope Science Institute, which is operated by AURA, Inc.,
under NASA contract NAS 5-26555 and by STScI grant GO-3594.01-91A}}
\author{A. Capetti\inst{1} \and H.R. de Ruiter\inst{2,3} 
\and R. Fanti\inst{4,3} \and R. Morganti\inst{5} \and P. Parma\inst{3}
\and M.-H. Ulrich\inst{6}
}
\institute{
Osservatorio Astronomico di Torino, Strada Osservatorio 25, 
I-10025 Pino Torinese, Italy
\and
Osservatorio Astronomico di Bologna, Via Ranzani, 1, I-40127 Italy
\and
Istituto di Radioastronomia, Via Gobetti 101, I-40129, Bologna, Italy
\and
Dipartimento di Fisica dell'Universit{\`a} di Bologna, Via Irnerio 46, 
I-40126 Bologna, Italy
\and
Netherlands Foundation for Research in Astronomy, Postbus 2,
NL-7990 AA, Dwingeloo, The Netherlands
\and 
European Southern Observatory, Karl-Schwarzschild-Strasse 2, D-85748,
Garching, Germany
}

\offprints{capetti@to.astro.it}

\date{Received 31 May 2000; accepted 31 August 2000}

\titlerunning{HST survey of the B2 sample}
\authorrunning{Capetti et al.}  \maketitle
\begin{abstract}

A Hubble Space Telescope snapshot survey of the B2 sample of low
luminosity radio galaxies has, at present, produced V and I images of 41
objects. Together with 16 images of B2 sources taken 
from the HST archive, there are now high resolution optical data 
for $\sim 57$~\% of the sample.
All host galaxies are luminous ellipticals, 
except one which is a spiral galaxy, while another one turns out to be a 
misidentification.

We present an album of the images of the B2 radio galaxies
observed so far, and give a brief description of the optical morphology 
of the galaxies.
Dust features (in the form of disks, lanes or irregular patches) are 
seen in most of the galaxies of the sample, $\sim 58$ \%. 
Compact optical cores are also very common (18/57). 
A preliminary analysis has revealed the presence of an optical
jet in three objects, indicating 
they can be detected in a sizeable percentage in 
these low luminosity radio sources.
Brightness profiles of dust-free galaxies are well represented by
a Nuker law and all shows the existence of a resolved shallow cusp.

\end{abstract}


\thesaurus{11.01.02,11.05.01,11.10.01,11.14.01,}

\section{INTRODUCTION}
\label{intro}

In the past few years, Hubble Space Telescope (HST) imaging of 
radiogalaxies has provided very valuable information on the optical 
structure of these sources. 
In particular, the HST/WFPC2 snapshot survey of
the 3CR sample produced a large and uniform database of images 
essential for a statistical analysis of their host galaxies (Martel et al. 
\cite{martel99}, De Koff et al. \cite{dekoff96}).

These studies revealed the presence of new and interesting features,
some of them almost exclusively associated to low luminosity 
FR~I radio-galaxies.

For example, HST observations have shown the presence of
dust in a large fraction of radio galaxies which, however,
takes the form of circum-nuclear disks (Jaffe et al. \cite{jaffe};
De Koff et al. \cite{dekoff96}; De Juan et al. \cite{dejuan}, 
Verdoes Klein et al.~\cite{verdoes}) only in FR~I sources. 
These structures have been naturally identified with the reservoir of
material which will ultimately accrete into the central black hole
and might provide a 
direct approach to measure key parameters such as the accretion rate 
and the black hole mass in AGNs. Although the precise relationship between
the symmetry axis of these disks and that of the sub-parsec scale
structure is not yet fully established, they can represent useful indicators
for the orientation of the central engine 
(Capetti \& Celotti \cite{capetti99}).
Several new optical jets have also been found in the HST images
(e.g. Sparks et al. \cite{sparks}, Baum et al. \cite{baum},
Martel et al. \cite{martel98}):
according to Martel et al. (\cite{martel99}) they are seen in 
$\sim 13$~\% of nearby ($z<0.1$) 3CR radio galaxies
and, with the only exception of 3C 273,
all are found in FR~I radio-galaxies.

FR~I sources also represent an essential ingredient in the Unified Schemes
of radio-loud AGNs as the mis-oriented counter part of 
BL Lac objects (see Urry \& Padovani \cite{urry95} for a review). 
Indeed, recent radio studies have
provided strong evidence in favour of
relativistic beaming in their jets (Laing et al. \cite{laing}).
Host galaxies of BL Lacs 
have been studied with the HST by Urry et al. (\cite{urry}),
and a similar analysis of galaxies hosting FR~I sources should clarify
if there is continuity in the properties of the respective nuclear 
regions, i.e. if any differences are due solely to orientation  effects.
Furthermore, Chiaberge, Capetti \& Celotti 
(\cite{chiaberge99}) have shown that
the nuclear sources commonly found in FR~I might represent the optical 
counter part of the synchrotron radio cores. 
This opens the possibility of testing
the unified models by directly comparing the anisotropic optical jet emission 
in BL Lac and FR~I. 

As many of the most extensively studied FR~I radio
galaxies are part of the B2 sample of low luminosity radio
galaxies (Fanti et al.~\cite{fanti87}),
this prompted us to perform a complete, high resolution optical study 
of these $\sim$ 100 sources. The radio 
characteristics of the B2 sample are different and in some sense 
complementary to those of the 3CR already studied with HST
and they can fill the gap between 
the "radio quiet \& normal"  and the "radio loud" ellipticals. 

A statistical study of such a large sample of low luminosity radio galaxies
will enable us to establish how frequently we can detect
optical jets, optical nuclear sources, circum-nuclear dusty disks 
and dust lanes and what is their relationship with the radio properties. 
Combining the HST observations of the B2 sample with 
those already existing of the complementary 3CR radio galaxies and 
BL Lac objects it will therefore be possible to find the similarities 
and differences of the optical nuclear properties as a function of 
radio power and morphological type of the parent optical objects. 
Particularly relevant will be, in this respect, the comparison of the optical 
brightness distribution of these samples of active sources with 
non active galaxies.

The organization of this paper is as follows; in 
Sect.~\ref{thesample} we briefly
describe the B2 sample and give some general information. 
In Sect.~\ref{hstobs} the present status of the HST observations is 
discussed and some relevant data on the program are given. 
The results are presented in Sect.~\ref{results}, in the form of 
images and notes on the individual sources. Finally, 
in Sect.~\ref{summary} we give a brief summary
and discuss future work planned for the B2 sample.  

Throughout this paper we use a Hubble constant 
H$_{\rm o}$= 100 km s$^{-1}$ Mpc$^{-1}$ and $q_0=0.5$.

\section{THE SAMPLE} \label{thesample} 

Here we give a short description of the properties of the B2 sample
of low luminosity radio galaxies. It consists of 
$\sim 100$ elliptical galaxies\footnote{the number of galaxies 
constituting the sample has recently undergone some changes. The 
present definition of 
the sample will be discussed elsewhere} identified with B2 radio sources
(Colla et al.~\cite{colla};  Fanti et al.~\cite{fanti78}).  The sample 
is complete down to 0.25 Jy at 408 MHz and down to a limiting 
magnitude 
$m_v$ = 16.5.  Since the sample was selected at low radio frequency,
it is largely unbiased for orientation. The sample spans the power range
between 10$^{22}$ and 10$^{25}$ WHz$^{-1}$ at 1.4 GHz with a pronounced peak 
around 10$^{24}$ WHz$^{-1}$ and therefore gives an excellent 
representation of the radio source types encountered below and around 
the break in the radio luminosity function. 

The B2 sample has been extensively 
studied at radio wavelengths, especially since the 1980's (see 
Fanti et al.~\cite{fanti87}, Parma et al.~\cite{parma}, De
Ruiter et al.~\cite{deruiter}, Morganti et al.~\cite{morg97}). 
Comparison with Einstein Satellite X-ray data was done by Morganti et 
al. (\cite{morg88}).  A number of B2 radio galaxies were subsequently 
observed with ROSAT (Feretti et al.~\cite{feretti},
Massaglia et al.~\cite{massaglia}, Trussoni et al.~\cite{trussoni}).  
Optical work on the sample has somewhat lagged behind, but
a complete broad band imaging survey of the B2 sample was carried out by 
Gonzalez-Serrano et al. (\cite{gonz93}) and Gonzalez-Serrano \&
Carballo (\cite{gonz00}),
while narrow band H$\alpha$ images were obtained by Morganti et al.
(\cite{morg92}). The IRAS properties of the sample were studied by 
Impey \& Gregorini (\cite{impey}). 

\section{OBSERVATIONS AND DATA REDUCTION}
\label{hstobs}

Up till the present HST imaging was done for 57 of the $\sim 100$ radio
galaxies. In the course of our HST program observations were obtained 
for 41 B2 galaxies, while public archive data exist for 16 additional 
objects, usually because these sources are also part of the 3C catalog 
(see De Koff et al. \cite{dekoff96}, Martel et al. \cite{martel99}, 
Verdoes-Klein et al.~\cite{verdoes}). The HST observations also confirmed
that the B2 radio galaxy 1441+26 is a misidentification 
(see Sect.~\ref{notes}) as already suggested by 
Gonzalez-Serrano \& Carballo (\cite{gonz00}).

Our program observations were obtained between April 9th and September 1st
1999, using the Wide Field and Planetary Camera 2 (WFPC2). 

The pixel size of the Planetary Camera,
in which all targets are located, is 0\farcs{0455} and
the 800 $\times$ 800 pixels cover a field of view of
$36^{\prime\prime} \times 36^{\prime\prime}$. 
Two broad band filters were used, namely F555W and F814W, which cover
the spectral regions 4500-6500 \AA\  and 7000-9500 \AA\  respectively.
Although their transmission curve does not match exactly those of
standard filters we will usually refer to them as V and I filters.
The data have been processed through the PODPS (Post Observation Data
Processing System) pipeline for bias removal and flat fielding
(Biretta et al. \cite{biretta}).

One image was taken through each filter and the exposure time was 
always set to 300 s (with the exception of B2~2116+26, which was 
observed for 350 s in the V band, as I band images were already 
available in the archive).

In order to minimize the overhead time associated to the read-out 
of the camera and to improve the efficiency of the program, 
we preferred to obtain a single image for each color
as this enables us, e.g., to study also the color distribution in the 
inner  regions of the galaxies. Obviously, with this choice, the 
removal of cosmic rays cannot be performed following the standard 
approach. 

We thus identified cosmic rays events based on the comparison
between the V and I images: 
each pixel where the ratio of the two images differed by more than a 
factor of 1.5 from its average value over the target and  
at the same time exceeded a threshold of 10 counts, was flagged as 
cosmic ray.
The region within an expansion radius of 1.5 pixel around each 
of them was replaced by interpolation of the nearest good pixels.
Clearly this method is appropriate only for extended targets free of
large gradients both in color and in intensity, such as are the objects
studied for this project. A small number of faint cosmic rays 
and hot pixels are still present in the cleaned images. 
These were removed individually (with the IRAF task IMEDIT). 
The overall fraction of bad pixels is $\sim 1$~\%: 
the images are not significantly affected by this process.
Visual inspection of these final images 
indicates that the resulting images can be considered to be essentially
free from strong cosmic rays that could have disturbed any further 
analysis (for example in the construction of brightness profiles). 

The conversion from counts to the standard WFPC2 photometric
system was derived using the header keyword PHOTFLAM of each filter
which is accurate to within 2 \% in the visible.

The archival
images were reduced following the standard PODPS and cosmic rays removal
procedure; the
observing log for these sources is reported in Tab. \ref{tab_archive}. 

\begin{table}
 \label{tab_archive}
\caption{Log of archival HST observations}
\begin{tabular}{l l c c c} \hline
Name         &  Filter & t$_{\rm exp}$ (s)& Date & Prog. ID             \\   \hline
0055+30      &  F814W  &            460   &   16/02/98 & 6673       \\
             &  F555W  &            460   &   16/02/98 & 6673           \\
0104+32      &  F702W  &            280   &   19/01/95 & 5476       \\     
0120+33      &  F555W  &           1600   &   13/07/97 & 6587       \\
1003+35      &  F702W  &            280   &   19/10/94 & 5476       \\
             &  F555W  &            600   &   12/06/96 & 6348       \\
1217+29      &  F814W  &            460   &   07/05/94 & 5454       \\
             &  F555W  &           1000   &   07/05/94 & 5454       \\
1251+27      &  F702W  &            560   &   16/03/94 & 5476       \\ 
             &  F555W  &            600   &   06/06/96 & 6348       \\
1321+31      &  F814W  &            460   &   04/11/98 & 6673       \\
             &  F555W  &            460   &   04/11/98 & 6673       \\
1350+31      &  F702W  &            280   &   15/01/95 & 5476       \\
1502+26      &  F702W  &            280   &   12/09/94 & 5476       \\
1511+26      &  F702W  &            280   &   29/11/94 & 5476       \\
1615+32      &  F702W  &            280   &   29/04/95 & 5476       \\
             &  F555W  &            600   &   19/10/96 & 6348       \\
1626+39      &  F702W  &            280   &   09/09/94 & 5476       \\
1726+31      &  F702W  &            280   &   25/01/95 & 5476       \\
1833+32      &  F702W  &            280   &   25/06/94 & 5476       \\
2229+39      &  F702W  &            280   &   06/08/94 & 5476       \\
2335+26      &  F702W  &            280   &   23/01/95 & 5476        \\
 \hline          				       
 \end{tabular}   
 \end{table}

The relevant data of the observed sources are 
given in Tab.~\ref{tabr}: in columns 1 and 2 we give the radio source
names (B2 and other), in column 3 the redshift, in columns 4 and 5 the
total and core power at 1.4~Ghz respectively, in columns 6 
and 7
the largest angular size in arcsec and the largest linear size in kpc,
the position angle of the main radio axis (in degrees) in column 8, 
where
a ``j'' indicates the direction of the radio jets; finally the 
Fanaroff-Riley 
type is given  in column 9. 
In a few cases no FR classification is possible (because of the
lack of sufficiently high resolution observations), while for some 
other sources the
FR type is given as I-II; this classification concerns sources with a 
hybrid structure, i.e. in which  both FR~I and FR~II characteristics 
are present.

\begin{table*}
\caption{Summary of the 1.4 GHz radio data of the sample.}\label{tabr}
\hspace{1.5cm} 
\scriptsize
\begin{flushleft}
\begin{tabular}{l l l c l r r r c} \hline\noalign{\smallskip}
B2   & Other & Redshift & $\log P_t$  & $\log P_c$    & l.a.s. & l.l.s. & p.a.  & FR \\
Name & Name  &          & (P in W/Hz) & (P in  W/Hz)  & arcsec & kpc    & degr. & type \\
\noalign{\smallskip}
\hline\noalign{\smallskip}
0034+25 &         & 0.0321 & 23.20 & 21.77  &  274 & 120 & -87\rlap{j} & I \\ 
0055+26 & 4C26.03 & 0.0472 & 24.61 & 22.30  &  240 & 150 & -40\rlap{j} & I \\
0055+30 &         & 0.0167 & 24.08 & 23.30  & 3450 & 811 & -50         & I  \\
0104+32 & 3C 31   & 0.0169 & 24.21 & 22.63  & 1900 & 452 & -20\rlap{j} & I \\
0116+31 & 4C31.04 & 0.0592 & 24.95 & 22.80  &  0\rlap{.1} & 0\rlap{.08} & -80 & \\
0120+33 &         & 0.0164 & 22.30 & 20.76  &  130 &  30 & -80 & I \\ 
0149+35 &         & 0.0160 & 22.33 & 21.54  &  191  &43 & 65 & I \\
0648+27 &         & 0.0409 & 23.62 & 22.92  & 0.02 & 0.011 & - & \\
0708+32 &         & 0.0672 & 23.51 & 23.02  &    8 &   7 &  -5 &  \\
0722+30 &         & 0.0191 & 22.67 & 22.02  &   35 &   9 & -85 & I \\
0755+37 &         & 0.0413 & 24.49 & 23.59  &  138 &  76 & -65 & I \\
0908+37 &         & 0.1040 & 24.84 & 23.46  &   51 &  63 &  15 & I-II \\
0915+32 &         & 0.0620 & 24.00 & 22.56  &  540 & 432 &  30 & I \\
0924+30 &         & 0.0266 & 23.52 & \llap{$<$}20.5  & 720 & 264 & 60 & I \\
1003+26 &         & 0.1165 & 24.01 & 22.10 &     6 &   8 &  10 &  \\
1003+35 & 3C236   & 0.0989 & 25.78 & 24.00 &  2400 & 2860 & -60 & II \\
1005+28 &         & 0.1476 & 24.25 & 22.60 &   240 & 391 & -25 & I-II \\
1101+38 &         & 0.0300 & 23.97 & 23.33 &   257 & 106 &  90 \\
1113+24 &         & 0.1021 & 23.65 & \llap{$<$}22.35  &  30 &  37 & 50  & I \\
1204+34 &         & 0.0788 & 24.47 & 23.05 &    51 &  50 & -50 & II \\
1217+29 &         & 0.0021 & 21.24 & 21.22 &  0\rlap{.03} & 0\rlap{.001} \\
1251+27 & 3C277.3 & 0.0857 & 25.37 & 23.30 &   45 &  48 & -22 & II \\
1256+28 &         & 0.0224 & 23.05 & 21.03 &  260 &  81 &     & I \\
1257+28 &         & 0.0239 & 23.08 & 20.69 &   20 &   7 &     & I \\
1321+31 &         & 0.0161 & 23.85 & 21.77 &  720 & 163 & -60 & I \\
1322+36 & 4C36.24 & 0.0175 & 24.55 & 22.38 &   53 &  13 &  15 & I \\
1339+26 & 4C26.41 & 0.0757 & 24.30 & \llap{$<$}22.5   & 285 & 271 & 30  & I \\
1346+26 & 4C26.42 & 0.0633 & 24.55 & 23.37 &   11 &   9 & 25  & I \\
1347+28 &         & 0.0724 & 24.05 & 22.27 &   47 &  43 & 50  & I-II \\
1350+31 & 3C293   & 0.0452 & 25.03 & 23.34 &  331 & 199 & 90\rlap{j} & I-II \\
1357+28 &         & 0.0629 & 24.03 & 22.45 &  139 & 113 & 0\rlap{j}  & I \\
1422+26 &         & 0.0370 & 24.00 & 22.24 &  140 &  70 & -85 &I \\
1430+25 & 4C25.46 & 0.0813 & 24.20 & \llap{$<$}21.9  & 50 &  51 & 20  & I \\
1447+27 &         & 0.0306 & 22.78 & 22.51 &    4 &   2 & 27 \\
1450+28 &         & 0.1265 & 24.50 & 23.01 &  200 & 290 & -60 & I \\
1455+28 & 4C28.38 & 0.1411 & 25.22 & \llap{$<$}23.03  & 213 & 336 & 38  & II \\
1457+29 &         & 0.1470 & 24.89 & 23.83 &  81 & 132 & -25 & I \\
1502+26 & 3C310   & 0.0540 & 25.36 & 23.40 &  260 & 184 & -15 & I \\
1511+26 & 3C315   & 0.1078 & 25.34 & \llap{$<$}24.3  & 125 & 160 &  & I \\
1512+30 &         & 0.0931 & 23.82 & 21.00 &   22 &  25 & -17 \\
1521+28 & 4C28.39 & 0.0825 & 24.58 & 23.50 &  200 & 205 & -30\rlap{j} & I \\
1525+29 &         & 0.0653 & 23.98 & 22.10 &   23 &  19 & 30\rlap{j} & I \\
1527+30 &         & 0.1143 & 24.05 & 22.80 &   45 &  60 & 45\rlap{j} & I \\
1553+24 &         & 0.0426 & 23.57 & 23.01 &  285 & 162 & -40\rlap{j} & I \\
1557+26 &         & 0.0442 & 22.81 & \llap{$<$}22.80  & 2 & 1 & - \\
1610+29 &         & 0.0313 & 22.93 & \llap{$<$}21.0  & 135 &  58 & 70  & I \\
1613+27 &         & 0.0647 & 24.03 & 22.69 &   31 &  26 & -37 & I \\
1615+32 & 3C332   & 0.1520 & 25.79 & 23.43 &   91 & 152 & 15  & II \\
1626+39 & 3C338   & 0.0303 & 24.49 & 23.16 &  103 &  43 & 90  & I \\
1658+30A & 4C30.31 & 0.0351 &23.88 & 22.89 &  160 &  76 & 55  & I-II \\
1726+31 & 3C357   & 0.1670 & 25.89 & 23.34 &  107 & 191 & -70 & II \\
1827+32 &         & 0.0659 & 24.07 & 23.08 &  360 & 304 & 75  & I \\
1833+32 & 3C382   & 0.0578 & 25.07 & 23.85 &  170 & 128 & 50  & II \\
2116+26 &         & 0.0164 & 22.79 & 22.08 &  457 & 106 & -23\rlap{j} & I \\
2229+39 & 3C449   & 0.0181 & 24.03 & 22.11 &  840 & 213 & 10\rlap{j} & I \\
2236+35 &         & 0.0277 & 23.47 & 21.74 &   47 &  19 & 45\rlap{j} & I \\
2335+26 & 3C465   & 0.0301 & 24.88 & 23.22 &  480 & 198 & -55\rlap{j} & I \\
\noalign{\smallskip}
\hline
\end{tabular}
\end{flushleft}
\end{table*}

\begin{figure} 
\vspace{8cm}
\includegraphics{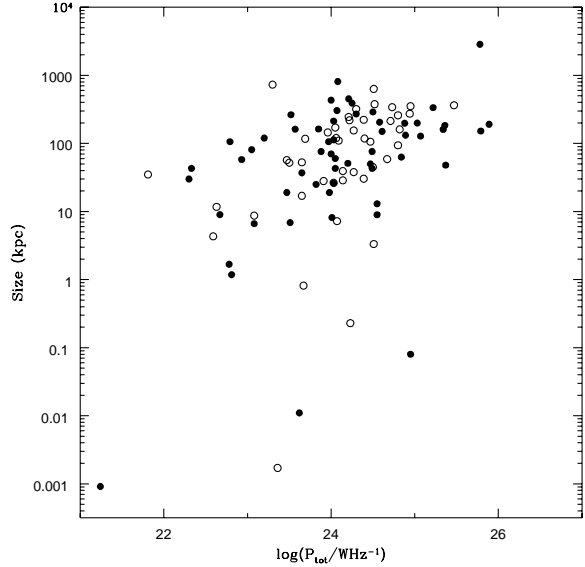} 
\caption{Size vs radio power diagram for sources observed (filled circles)
and not yet observed (empty circles) with HST} 
\label{fig:pd} 
\end{figure}

\begin{figure} 
\vspace{8cm}
\includegraphics{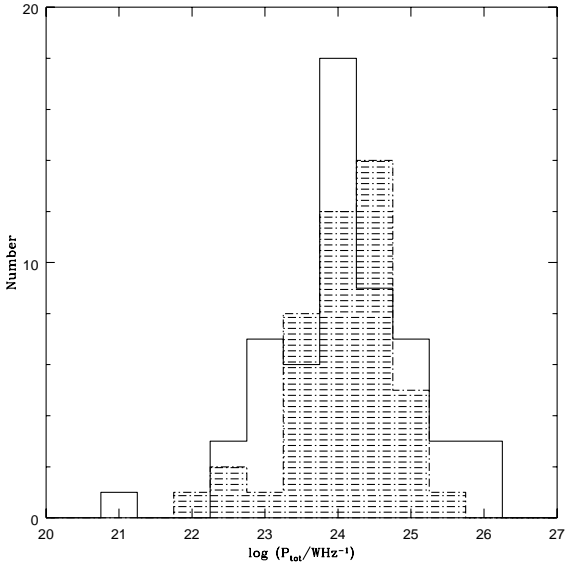} 
\caption{Distribution of total power for sources observed
and not yet observed (shaded) with HST} 
\label{fig:pthist} 
\end{figure}

\begin{figure} 
\vspace{8cm} 
\includegraphics{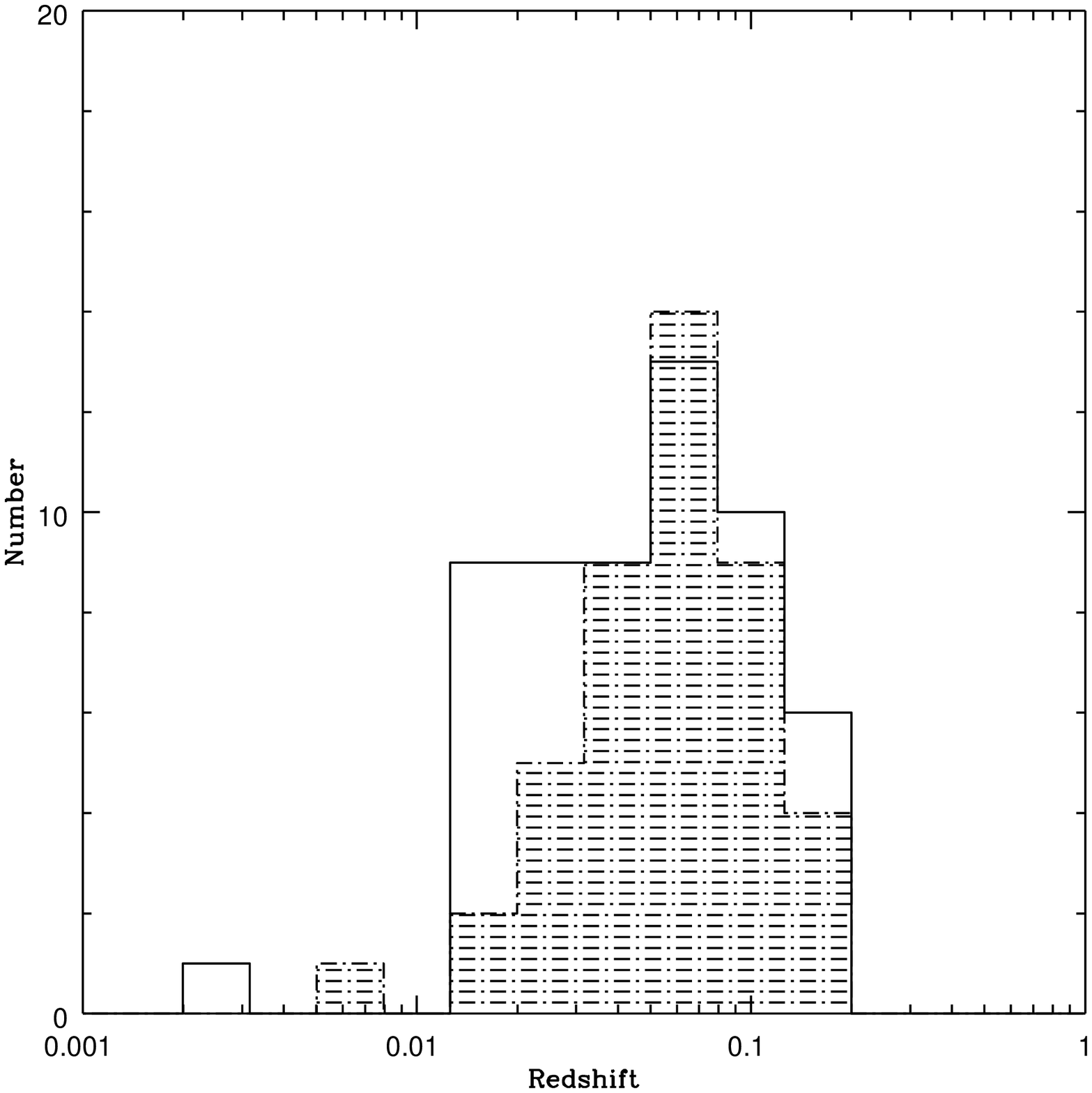} 
\caption{Distribution of
redshifts for sources observed and not yet observed (shaded) with HST}
\label{fig:zhist} 
\end{figure}

While there is no bias in the selection of the sources  of our program
(they were chosen randomly as far as their radio and optical properties are 
concerned), a bias might have been introduced by the inclusion 
of the 16 archive sources. 
We thus compared various parameters 
of the observed sub-sample with those of the sources
that were not observed by HST.
This is illustrated in
Figs. \ref{fig:pd}, ~\ref{fig:pthist} and \ref{fig:zhist};
no significant differences emerged except for
a small effect in redshift and total power.  
The median redshift and radio power
of the two sub-samples are $ z = 0.055^{+0.008}_{-0.011}$,
log $P_t = 24.05^{+0.07}_{-0.03}$ and
$ z = 0.067^{+0.005}_{-0.007}$,
log $P_t = 24.22^{+0.06}_{-0.03}$ respectively, 
thus only marginally different.
We can then consider the sources observed with HST as 
well representative of the whole B2 sample.  

\section{RESULTS}
\label{results}

In Fig. 4 we present the final broad
band images of the innermost regions of all 57 B2 radio galaxies
observed. In the left (right) bottom corner is indicated the size of the 
field of view in arcsec (kpc).
Since the B2 sample contains $\sim 100$ objects this means
that at present the HST observation are complete at the level of 
$\sim 57$~\%.

A preliminary analysis of the HST images is presented in the notes on
individual sources given below. In particular we checked for the 
presence of dust features, a compact optical core and an optical jet.
The identification of optical cores is based on the procedure
described in Chiaberge et al. (\cite{chiaberge99}),  
i.e. a source brightness profile which, within 5 pixels from the center, 
shows a FWHM consistent with the HST Point Spread Function 
($\le 0\farcs08$). 
As to the presence of optical jets, we just checked 
that after subtraction of a model galaxy, some emission was left at 
the location of the radio jet. 
We stress that a more detailed analysis will be carried out by us 
elsewhere.
A summary of the optical data is given 
in Tab.~\ref{tabo}, in which we give the names of the sources (B2 and 
other) in columns 1 and 2, the redshift in column 3; the presence of 
dust is indicated in column 4 as d (disk-like structure), l (dust lane, 
band or filament), p (irregular patches); if possible we measure a 
position angle of the disk or band and the P.A. (in degrees from north 
through east) is given in column 5; the presence of unresolved optical 
cores and jets is indicated in columns 6 and 7. 

\subsection{Notes on the individual sources}
\label{notes}

\begin{table} 
\caption{Summary of optical data of the sample.}\label{tabo}
\hspace{1.5cm} 
\scriptsize
\begin{flushleft}
\begin{tabular}{l l l c r c c} \hline\noalign{\smallskip}
 Name & Other  & Redshift & \multicolumn{2}{c}{Dust} & Core & Jet \\
      & Name   &          & Morph.  & P.A. & Opt. & Opt. \\
\noalign{\smallskip}
\hline\noalign{\smallskip}
0034+25 & UGC00367  & 0.0321   & d   & 160 &     &    \\
0055+26 & NGC0326   & 0.0472   &     &     &     &    \\
0055+30 & NGC0315   & 0.0167   & d+p &  40 & yes &    \\
0104+32 & NGC0383   & 0.0169   & d   &  40 & yes &    \\
0116+31 &           & 0.0592   & l   &  25 &     &    \\
0120+33 & NGC0507   & 0.0164   &     &     &     &    \\
0149+35 & NGC0708   & 0.0160   & l+p &     &     &    \\
0648+27 &           & 0.0409   & p   &     &     &    \\
0708+32 &           & 0.0672   &     &     &     &    \\
0722+30 &           & 0.0191   & l   &     &     &    \\
0755+37 & NGC2484   & 0.0413   &     &     & yes & yes \\
0908+37 &           & 0.1040   & d?  &  40 & yes &    \\
0915+32 &           & 0.0620   & d   &     &     &    \\
0924+30 &           & 0.0266   &     &     &     &    \\
1003+26 &           & 0.1165   &     &     &     &    \\
1003+35 &           & 0.0989   & l   &  50 &     &    \\
1005+28 &           & 0.1476   & p   &     &     &    \\
1101+38 & MRK 421   & 0.0300   &     &     & yes &    \\
1113+24 &           & 0.1021   &     &     &     &    \\
1204+34 &           & 0.0788   &     &     &     &    \\
1217+29 & NGC4278   & 0.0021   &     &     & yes &    \\
1251+27 & Coma A    & 0.0857   & p   &     & yes &    \\
1256+28 & NGC4869   & 0.0224   & d?  &     &     &    \\
1257+28 & NGC4874   & 0.0239   &     &     &     &    \\
1321+31 & NGC5127   & 0.0161   & l   &  45 &     &    \\
1322+36 & NGC5141   & 0.0175   & l   &  85 &     &    \\
1339+26 & UGC08669  & 0.0757   & l   &   0 &     &    \\
1346+26 &           & 0.0633   & l   &   0 & yes &    \\
1347+28 &           & 0.0724   & p   &     &     &    \\
1350+31 & UGC08782  & 0.0452   & l+p &     &     &    \\
1357+28 &           & 0.0629   & l   &  95 &     &    \\
1422+26 &           & 0.0370   &     &     &     &    \\
1430+25 &           & 0.0813   &     &     &     &    \\
1447+27 &           & 0.0306   & l   & 145 &     &    \\
1450+28 &           & 0.1265   &     &     &     &    \\
1455+28 &           & 0.1411   & l+p &     &     &    \\
1457+29 &           & 0.1470   & l   &  30 &     &    \\
1502+26 &           & 0.0540   &     &     & yes &    \\
1511+26 &           & 0.1078   &     &     &     &    \\
1512+30 &           & 0.0931   & p   &     &     &    \\
1521+28 &           & 0.0825   &     &     & yes &    \\
1525+29 & UGC09861  & 0.0653   & l   &  25 &     &    \\
1527+30 &           & 0.1143   & d?  &     &     &    \\
1553+24 &           & 0.0426   &     &     & yes & yes \\
1557+26 &           & 0.0442   &     &     &     &    \\
1610+29 & NGC6086   & 0.0313   &     &     &     &    \\
1613+27 &           & 0.0647   & p   &     &     &    \\
1615+32 &           & 0.1520   &     &     & yes &    \\
1626+39 & NGC6166   & 0.0303   & p   &     & yes &    \\
1658+30A&           & 0.0351   &     &     &     & yes \\
1726+31 &           & 0.1670   & l   &     &     &    \\
1827+32 &           & 0.0659   &     &     &     &    \\
1833+32 &           & 0.0578   &     &     & yes &    \\
2116+26 & NGC7052   & 0.0164   & d   &  65 & yes &    \\
2229+39 & UGC12064  & 0.0181   & l   & 165 & yes &    \\
2236+35 & UGC12127  & 0.0277   &     &     & yes &    \\
2335+26 & NGC7720   & 0.0301   & d   &     & yes &    \\
\noalign{\smallskip}
\hline
\end{tabular}
\end{flushleft}
\end{table}

{\bf 0034+25:}
There is a prominent dust band in the center of this galaxy, likely a 
highly inclined
dusty disk oriented along P.A. $\sim 160^\circ$. At larger radii,
isophotes are very cuspy and oriented at the same position angle as the 
inner disk, which indicates the presence of a stellar disk coplanar 
with the inner one. The galaxy belongs to the Zw cluster 0034.4+2532.

{\bf 0055+26 (NGC~326):}
This is a system of two galaxies, also referred to as a dumbbell galaxy.
Only the galaxy hosting the radio source is shown. The two galaxies do 
not show any peculiarities. There is a third much fainter object in 
the field. The galaxy belongs to a group with extended X-ray emission 
(Worrall et al.~\cite{worrall})

{\bf 0055+30 (NGC~315):}
A highly inclined, very regular circum-nuclear disk, is seen in absorption
in this galaxy. It is oriented at P.A. $\sim 40^\circ$ and extends to
$r \sim 350$ pc. At its center there is an unresolved nuclear source.
Irregular dust patches are also present on the South side of the galaxy.
See also Verdoes Kleijn et al. (\cite{verdoes}).

{\bf 0104+32 (NGC~383, 3C~31):} An extended circum-nuclear dusty disk is 
seen at low inclination, with an unresolved nuclear component
at its center. The disk diameter is $\approx 7 \arcsec\ $
(2.5 kpc) along the major axis (P.A. $\sim 40^\circ$). The galaxy 
belongs to a chain of which it is the brightest member.
See also Martel et al. (\cite{martel99}) and Verdoes Kleijn et al. 
(\cite{verdoes}).

{\bf 0116+31:}
An S-shaped dust lane bisects its nuclear region. Its inner P.A. is 
$\sim 25^\circ$. Conway (\cite{conway}) detected a disk of HI in 
absorption which fully occults one of the minilobes and partially 
covers the other.

{\bf 0120+33 (NGC~507):}
Brightest galaxy of the Zw cluster 0107+3212. The galaxy appears 
very regular and smooth.

{\bf 0149+35 (NGC~708):}
A low brightness galaxy whose nuclear regions are crossed by an
irregular dust lane and dust patches.
It is the brightest galaxy of the cluster Abell 262. 

{\bf 0648+27:}
A galaxy with an irregular morphology, enhanced by the presence
of dusty features. On its northern side there is a chain of knots,
probably regions of star formation, parallel to the dusty filaments.

{\bf 0708+32:}
The host galaxy does not exhibit any obvious peculiarities.

{\bf 0722+30:}
The only source of the B2 sample associated with a spiral galaxy.
It is a highly inclined galaxy which fills the PC field of view.
Strong absorption is associated with its disk and a bulge-like 
component is also clearly visible. The radio emission originates from 
two symmetric jet-like features at an angle of $45^{\rm o}$ to the disk.

{\bf 0755+37 (NGC~2484):}
The galaxy is characterized by a central unresolved nuclear source.
After subtraction of a galaxy elliptical model the presence of 
a one-sided optical jet is revealed, cospatial with the SE brighter 
radio jet.

{\bf 0908+37:}
This galaxy shows a point-like nucleus at the center of a highly 
elongated absorption feature, possibly an edge-on disk 
(P.A. $\sim 40^\circ$).
A fainter companion galaxy (outside the image shown) is located 
$ 3\farcs7$\ to the SW.

{\bf 0915+32:}
A disk-like dust band surrounds its nuclear region.

{\bf 0924+30:}
The host galaxy does not exhibit any obvious peculiarities.

{\bf 1003+26:}
A very regular low brightness elliptical galaxy
with no outstanding morphological features. The galaxy belongs to the 
cluster Abell 923.

{\bf 1003+35 (3C~236):} An off-centered dust lane crosses the galaxy to 
the SW,
with $\approx 6 \arcsec\ $ (2.5 kpc) extension. See also 
Martel et al. (\cite{martel99}). 

{\bf 1005+28:}
A diffuse and quite regular galaxy with no evident morphological 
peculiarities. A dust patch is visible.

{\bf 1101+38:}
This source is associated with the BL Lac object Mrk 421.
Not surprisingly it is dominated by its bright unresolved nucleus
which produces strong diffraction spikes. Nonetheless a smooth 
elliptical host galaxy is clearly visible.

{\bf 1113+24:}
A round smooth elliptical galaxy with a faint companion 
$\sim 3\arcsec\ $ to the north. The galaxy is the brightest of the  
Zw cluster 1113.0+2452.

{\bf 1204+34:}
Dust patches produce a spiral structure, which is clearly visible 
in both bands. 
It extends out to at least 2\arcsec\ from a fully resolved 
nucleus. A secondary, again resolved,  nucleus 
is found to the NE at a distance of $\approx 2\arcsec\ $. 
Both nuclei are embedded in a common elliptical halo, centered on the
brighter one.

{\bf 1217+29 (NGC~4278):}
This source is associated with the nearby galaxy NGC 4278.
It shows a regular structure only slightly modified by diffuse dust 
absorption to its NW side. The nucleus is unresolved. The radio source 
is very compact, $\approx 1$ pc (see, e.g., Schilizzi et 
al.~\cite{schilizzi}). The galaxy is known to have strong nuclear 
emission lines (Osterbrock~\cite{osterbrock}). It contains a large 
amount of neutral hydrogen (Raimond et al.~\cite{raimond}) rotating 
around an axis at $-45 ^\circ$. 

{\bf 1251+27 (Coma A, 3C~277.3):}
In addition to a prominent unresolved nucleus, note also the 
filamentary structure located $\approx 7 \arcsec\ $ to the SW. This 
corresponds to the  radio knot K1 in van Breugel et al. 
(\cite{breugel}), detected also in emission lines.
See also Martel et al. (\cite{martel99}). 

{\bf 1256+28 (NGC~4869):}
This galaxy presents a highly elongated
absorption feature, possibly an edge on disk. It belongs to the Coma 
Cluster (Abell 1656).

{\bf 1257+28 (NGC~4874):}
A very regular low brightness elliptical galaxy. 
It is one of the two dominant members of the Coma Cluster
(Abell 1656).
 
{\bf 1321+31 (NGC~5127):}
A dark lane covers the nucleus of this galaxy and extends out to a 
radius of $ 1\arcsec\ $, P.A. $\sim 45 ^\circ$. The galaxy belongs to 
the Zwicky cluster 1319.6+3135. See also Verdoes Kleijn et al. 
(\cite{verdoes}).

{\bf 1322+36 (NGC~5141):}
There is a wide dust band in the center of this galaxy, at 
P.A. $\sim 85^\circ$. See also Verdoes Kleijn et al. (\cite{verdoes}).

{\bf 1339+26:}
A small dark nuclear band  (P.A. $\sim 0^\circ$) 
characterizes this otherwise very regular galaxy.
The galaxy is the eastern of a double system and is the dominant 
member of the cluster Abell 1775.

{\bf 1346+26:}
A low brightness galaxy with an unresolved nucleus 
and dust in the form of an irregular lane and patches 
(P.A. $\sim 0 ^ \circ$). 
A sort of tail is seen to the SW towards a companion.
It is the brightest galaxy (cD) of the cluster Abell 1795.

{\bf 1347+28:}
A regular elliptical galaxy with irregular dusty features. It belongs 
to Abell 1800.
 
{\bf 1350+31:}
A highly irregular galaxy, with two compact emission knots separated by 
a filamentary dust lane which extends to form a fan-like shape at 
larger radii. See also Martel et al. (\cite{martel99}). See  
Akujor et al. (\cite{akujor}) for a high resolution
radio map showing a flat spectrum core with a two sided jet.

{\bf 1357+28:}
A round elliptical galaxy with a small dust lane (P.A. $\sim 95 ^ \circ$) 
bisecting the nuclear region.

{\bf 1422+26:}
The host galaxy is very smooth with no outstanding morphological 
features. The bright central core is fully resolved.

{\bf 1430+25:}
This elliptical galaxy does not exhibit any obvious peculiarities. An 
elongated nucleus has its major axis at P.A. $\sim 155 ^ \circ$. 
The radio-structure is head-tail type, but no obvious cluster or group 
is visible.

{\bf 1441+26:}
The target is a spiral galaxy. However, comparison of
the radio and optical maps shows that this is not the host of the 
radio source. The correct identification is with a faint elliptical 
galaxy located  at RA = 14:41:56.35, Dec = 26:14:05.0.  As the 
magnitude of this galaxy is well below the optical selection 
threshold of the B2 sample, it must be considered a misidentification.

{\bf 1447+27:}
Two off-center linear dust lanes run 0\farcs6 and 1\farcs2 respectively
north-east of the nucleus (P.A. $\sim 145 ^ \circ$) 
of this otherwise very regular galaxy. 

{\bf 1450+28:}
Elliptical galaxy with no outstanding features. In Abell 1984.

{\bf 1455+28:}
The presence of hour-glass shaped dust absorption 
gives this galaxy a quite irregular appearance. The central knot
is compact but completely resolved. Companion galaxy at $\approx
10\arcsec\ $. 

{\bf 1457+29:}
A wide dust lane (P.A. $30 ^ \circ$) runs perpendicularly to the galaxy 
projected minor axis. The redshift of the galaxy has been measured
recently by Gonzalez-Serrano \& Carballo (\cite{gonz00}); this
value has been added in Tab.~\ref{tabr}.

{\bf 1502+26 (3C~310):}
A smooth elliptical galaxy with a central unresolved source.
See Martel et al. (\cite{martel99}).

{\bf 1511+26 (3C~315):}
This galaxy shows a very high nuclear ellipticity
which contrasts with the typical roundness of the B2 host galaxies.
The ellipticity decreases at larger radii but 
the central very elongated central structure clearly 
extends at larger radii, probably indicative of a edge-on stellar disk. 
The radio source has a very peculiar, two-banana shaped, 
radio-structure. See de Koff et al. (\cite{dekoff96}).

{\bf 1512+30:}
The host galaxy does not show any outstanding morphological features, 
except for very faint elongated dust absorption at the center. 

{\bf 1521+28:}
A well defined unresolved nucleus is superposed on the smooth core of 
the galaxy.
 
{\bf 1525+29:}
The double peaked nuclear structure is caused by dust absorption, which
is extended in the direction P.A. $\sim 25 ^ \circ$. The galaxy is 
the fainter of a pair in the cluster Abell 2079.

{\bf 1527+30:}
The faint absorption from a possibly disk-like dust band is visible on the 
west side of this galaxy, extending to a radius of 0\farcs3.
Its central region is resolved. In Abell 2083.

{\bf 1553+24:}
This galaxy exhibits a point-like nucleus and, after removal of an 
elliptical model for the galaxy emission, a faint one-sided optical 
jet, cospatial with the NW brighter radio jet.

{\bf 1557+26:}
A smooth and regular elliptical galaxy.

{\bf 1610+29 (NGC~6086):}
This galaxy does not exhibit any obvious peculiarities. It is the 
brightest galaxy of the cluster Abell 2162.

{\bf 1613+27:}
Very faint absorption features are seen to the SW on this otherwise
regular galaxy.

{\bf 1615+32 (3C~332):}
The emission of this source is dominated by its bright unresolved 
nucleus. A companion is seen to SW.

{\bf 1626+39 (NGC~6166, 3C~338):}
Brightest galaxy of the cluster Abell 2199. An unresolved nucleus is 
superposed on the flat core of 
the galaxy. An arc-like dust feature extends from to nucleus to the W
for $\approx 3 \arcsec\ $. Two fainter elliptical companions are also 
present. (see Martel et al.~\cite{martel99}).

{\bf 1658+30:}
Smooth galaxy with a central unresolved core.
The subtraction of an elliptical galaxy model reveals the presence
of a faint one-sided optical jet, cospatial with the radio jet.

{\bf 1726+31 (3C~357):}
The central knot is well resolved. A dust lane is also present on 
its SW side.

{\bf 1827+32:}
Round galaxy with a possible weak nucleus.

{\bf 1833+32 (3C~382):}
This source is associated with the broad line radio galaxy 3C382. 
The nuclear emission
dominates the optical emission but the underlying host galaxy is 
clearly seen in these images. See also Martel et al. (\cite{martel99}).

{\bf 2116+26 (NGC~7052):}
A spectacular, highly inclined disk-like structure of dust 
(at P.A. $\sim 65^\circ$) surrounds 
the central source of this galaxy. See also Verdoes Kleijn et al. 
(\cite{verdoes}).
From the kinematics of the gas disk
van der Marel \& van den Bosch (\cite{marel}) deduced the presence of 
a central black hole with a mass of $3 \times 10^8 M_\odot$.

{\bf 2229+39 (3C~449):}
A wide band of absorption is located on the West side of this source,
at P.A. $\sim 165^\circ$.
See also Martel et al. (\cite{martel99}).
Note also the prominent unresolved nuclear source.

{\bf 2236+35:}
A faint nucleus, probably unresolved, is seen in this
elliptical galaxy.

{\bf 2335+26 (NGC~7720, 3C~465):}
It is the brightest galaxy in the cluster Abell 2634.
An elliptical region of dust absorption is located around the bright 
unresolved nucleus of this galaxy.
See also Martel et al. (\cite{martel99}).

\subsection{Discussion}
\label{discussion}

All B2 sources are hosted by bright elliptical galaxies, with the only
exception of B2~0722+30, which is associated to a spiral galaxy.
While this result might appear surprising, we must note that
both the size (9 kpc) and the radio luminosity ($\log P_t = 22.67$)
of this source place it at the lowest end of the range covered by the B2 
sample. On the other hand these values are quite typical of 
radio-sources associated to Seyfert galaxies (Nagar et al. \cite{nagar}) 
which are indeed associated to spiral galaxies.

Brightness profiles of all elliptical host galaxies which are not severely 
affected by dust 
extinction can be successfully represented by a Nuker law 
(Lauer et al. \cite{lauer}).
In all cases we found evidence for a well resolved, shallow 
($\gamma \le 0.3$) central cusp,
as expected given their range of absolute magnitudes,
$M_V$ from $-20$ to $ -23$.

In 18 of the 57 ( $\sim 32$~\%) 
B2 sources observed with the HST we found evidence 
for an unresolved nuclear source 
(see Tab.~\ref{tabo}) down a magnitude level of V = 23.5. The detection rate is
slightly lower than the percentage (between 43 and 54 \%) found
for 3C sources with $z<0.1$ (Martel et al. \cite{martel99}).
This discrepancy can be accounted for noting that B2 sources have on average
fainter radio cores than 3C sources (Giovannini et al. 
\cite{giovannini}) and that
the optical emission in FR~I nuclei strongly correlate with the radio 
core flux 
(Chiaberge et al. \cite{chiaberge99}). We thus expect that 
B2 galaxies harbour fainter optical cores
which might be more difficult to detect against the galaxy background.
Indeed the B2 sources in which we detected an optical core have higher radio
core fluxes and more dominant cores than average.

Dust is frequently present in B2 sources: after visual inspection of the images
we found that 33/57 (58~\%) of them show dust features,
either in the form of bands or disk-like structures, or more
irregular patches. The percentage may be slightly higher than in
3C sources with $z<0.1$, which have 37-48~\% of sources with dust.
Concerning the presence of dust features it does not appear that
there is any difference between
FR I and FR~II objects: the FR~I sources with dust are 19/35 (54~\%),
while the FR~I-II and FR~II with dust are 8/12 (66~\%). 
Our data strengthen the findings by  Martel et al. (\cite{martel99})  
which found disk-like structures 
only associated to FR~I type radio sources
(see Tabs.~\ref{tabr} and \ref{tabo}). Indeed in the B2 sample,
although dust is equally present in FR~I and FR~II sources, 
it is found to be ordered into disk-like structures in 9 FR~Is while 
this behaviour is never found in FR~IIs. 
However, the detection of dusty disks is almost exclusive of
nearby (z $\lta$ 0.03) galaxies. 
Within this distance there are only a handful
of FR~II in the 3C sample and none in the B2. This raises the question
of the importance of the observational biases in our ability of
finding such disks in FR~II sources. 
Among the 18 sources with an optical core 9 also show 
the presence of 
dust, while no dust is detected in the remaining 9; a slightly higher
rate of dust occurrence (24/39) is found among the sources
without optical core. 
This suggests that extended dust structures might be hiding optical 
cores in a few cases.
We find no connection between dust content and with largest radio linear size: 
the sources with dust have a median
linear size of $94\pm 35$ kpc, against $76\pm 30$ for the sources 
without dust. 

A first scrutiny of the B2 data
indicates that detection of optical counter part of radio jet
is not particularly rare. At present
we have found three optical jets, in B2~0755+37, B2~1553+24 and
B2~1658+30A (see Tab.~\ref{tabo}), but we do not exclude that a more
refined analysis will produce a few more. Therefore optical jets in 
HST images of 
these low luminosity radio galaxies might be detected in a sizeable 
fraction of the sample,
probably comparable with the 13 \% found in the low redshift 3C sources.
All optical jets are associated to
strongly one-sided sources, of relatively small linear sizes and with 
prominent radio nuclei in agreement with the results 
by Sparks et al. (\cite{sparks}).

\section{SUMMARY AND FUTURE WORK}
\label{summary}

We have presented HST/WFPC2 snapshot images in two bands (V and I)
of 57 (out of a total of $\sim 100$) sources
from the B2 sample of low luminosity radio-galaxies.
One more object (1441+26), not shown in Fig. 4 has been 
found to be a mis-identification 
since the radio source appears to be associated with
a background galaxy whose magnitude is below the optical 
selection threshold of the sample. 
Except for B2~0722+30, which is associated with a spiral galaxy,
all sources are hosted by bright elliptical galaxies well fitted
by Nuker laws, with shallow central cusps.

Most of the galaxies present a peculiar morphology. Almost 60~\% 
show the presence of large scale dust structures which in 
9 cases (all FR~I radio sources) is distributed in well defined disks.  
In three sources we detected an optical synchrotron jet.
The presence of central unresolved sources is revealed in 18 galaxies.

The present work is only a first presentation of the HST snapshot 
survey of B2 radio galaxies. While awaiting completion of the survey, 
more detailed studies will be presented in the near future. In 
particular an analysis of the brightness profiles will be given 
with a comparison
of samples of powerful radio galaxies as well as non active galaxies,
together with an exhaustive search for optical jets. Because of the 
relatively large number of objects it may be possible to study the 
orientation of optical features, such as dust-disks and -lanes,
as compared to the main radio axes.
Also the statistics of unresolved cores will be discussed in the
framework of Unified Models, and in this respect a comparison 
with the sample of BL Lac objects that has already been observed with 
the HST will become crucial.

\begin{acknowledgements}
This research has made use of the NASA/IPAC Extragalactic Database (NED)
which is operated by the Jet Propulsion Laboratory, California Institute of
Technology, under contract with the National Aeronautics and Space
Administration. 

\end{acknowledgements}

\end{document}